\documentclass[a4paper,12pt]{article}

\usepackage[latin1]{inputenc} %
\usepackage[T1]{fontenc} %
\usepackage{RR}
\usepackage{amsmath}    
\usepackage{graphicx}   
\usepackage{verbatim}   
\usepackage{color}      
\usepackage{subfigure}  
\usepackage{hyperref}   
\usepackage{delarray}

\newcommand{\AsiT}{A_{si}^T}
\newcommand{\AdjT}{A_{dj}^T}
\newcommand{\AsiF}{A_{si}^F}
\newcommand{\AdjF}{A_{dj}^F}
\newcommand{\Fsi}{F_{si}}
\newcommand{\Fdj}{F_{dj}}
\newcommand{\fij}{f_{ij}}
\newcommand{\pij}{p_{ij}}
\renewcommand{\Psi}{\phi_{si}}
\newcommand{\Pdj}{\phi_{dj}}
\newcommand{\fijSCRIP}{\fij^{\rm{SCRIP}}}
\newcommand{\fijINT}{\fij^{\rm{SCRIP,int}}}
\newcommand{\beq}{\begin{equation}}
\newcommand{\eeq}{\end{equation}}

\newtheorem{descr}{Description}
\newtheorem{hypothesis}{Hypothesis}
\newtheorem{definition}[descr]{Definition}

\RRdate{February 2013}

\RRNo{0001}
\RRauthor{
Jo\"el Chavas \and
\'Edouard Audit \and
Laure Coquart \and
Sophie Valcke
}
\authorhead{Chavas, Audit, Coquart \& Valcke}

\RCMDLS 

\RRetitle{Conservative Regridding When Grid Cell Edges Are Unknown -- Case of SCRIP}
\titlehead{Conservative regridding}
\RRabstract{Nowadays, climate models rely on couplers. Each complete climate model is broken into different sub-models (oceanic, atmospheric,$\dots$), each one working on a different grid. The coupler brings these models together and interpolates the physical quantities between the grids. However, neither the coupler nor sometimes the sub-models themselves know precisely the grid cell edges. They only know the grid cell corners (vertices) and the true grid cell areas. Thus, the coupler has to make assumptions about the grid cell edges in order to compute the grid cell intersections. For first-order schemes, the most straightforward way to interpolate scalar quantities is to directly use these approximate grid cell intersections, that don' take the true grid cell areas into account. It is the method used in the ''conservative'' regridding option implemented in the widely used spherical interpolation package SCRIP. We show that is doesn't preserve integrals in the general case, whether the coupler using the 
SCRIP-generated 
weights transmits directly the intensive quantity $F_{si}$, or its extensive counterpart, $F_{si}A_{si}^T$ (that is the same quantity multiplied by the true area of the individual source grid cell). We show how to modify the interpolation scheme to preserve integrals in the general case.}
\RRkeyword{conservative regridding, coupler, SCRIP, Lagrange multiplier, climate}

\begin{document}
\makeRT   


\tableofcontents
 ~ 
 \vfill
 \newpage

\section{Introduction}

SCRIP is a library widely used by different couplers between oceanic and atmospheric models~\cite{_scripusers.pdf_, jones_first-_1999}. In particular, it implements an option, the ''conservative'' option, that should let the coupler preserve integrals of a quantity to be mapped between the two grids. 

\begin{figure}[htb]\label{fig:regridding}
  \centering
  \def\svgwidth{\linewidth}
  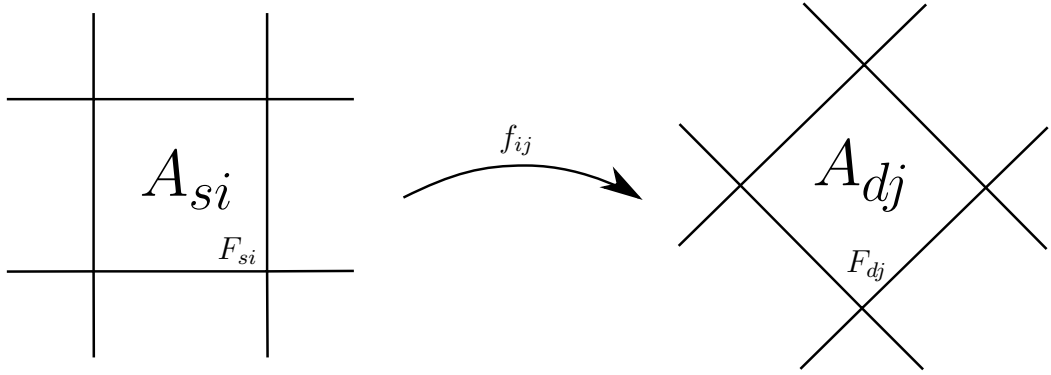
  \caption{Scheme of the regridding~: we interpolate from a source grid (underscript s) to a destination grid (underscript d). We note $A_{si}$, the area of the i$^{th}$ cell of the source grid. We note $A_{dj}$, the area of the j$^{th}$ cell of the destination grid. From $A_{si}$, $A_{dj}$ and the grid cell corners (vertices), we determine the interpolation weights $\fij$. We treat first-order schemes, so the intensive quantity $\Fsi$ is constant over the i$^{th}$ source grid cell. The interpolated quantity $\Fdj$ is constant over the j$^{th}$ destination grid cell. It reads $\Fdj = \sum_i\fij\Fsi$.}
\end{figure}

SCRIP permits to map and interpolate a quantity F between two numerical grids in spherical coordinates (Fig.~\ref{fig:regridding}). The only information given to SCRIP are the vertices of each source and destination grid cell. SCRIP, not knowing the true area of each grid cell, has to estimate these areas based on the coordinates of the vertices.  As such estimated areas can differ from the true ones, the conservative option of SCRIP is mathematically wrong in the general case. 

Recently, different reports in the climate community have addressed this issue~\cite{_regridding_karletaylor_v7_b.pdf_, _work_report_}. K.E.Taylor also suggested that the SCRIP library could be conservative in the general case, provided that the coupler multiplies the quantity to remap by the true area of the grid cell.\cite{_email_taylor_}

We will show under which conditions SCRIP is indeed conservative. We will give the minimum modifications that should be brought to make it conservative in the general case.

Apart from the use of the word 'SCRIP', which is the name of the interpolation library used in climate, all the mathematics of this report are general. It applies whenever we interpolate fields between two grids that have unknown grid cell edges.

\Large
\normalsize

\section{Scope of the report}

We want to know how the integral of a quantity can be preserved when interpolating between two grids. By hypothesis, the source map ''contains'' the destination map (see \S\ref{par:hypotheses}). So, we don't analyze the issues linked to non-matching sea-land mask.

\section{Notations}

The quantity to be regridded is called F. F is an intensive quantity, such as the temperature. F is mapped between a source grid S and a destination grid D (Fig.~\ref{fig:regridding}). All the quantities linked to the source grid take the underscript s ($X_s$), whereas all the quantities linked to the destination grid take the underscript d ($X_d$). The underscript i ($X_i$) always means ''a given cell of the source grid'', whereas the underscript j ($X_j$) always means ''a given cell of the destination grid''. The superscript T ($X^T$) stands for true, whereas the superscript F ($X^F$) stands for false. The letter A stands for area. $A_1\cap A_2$ stands for the area of the intersection of the two corresponding surfaces $S_1$ and $S_2$. 

Thus,
\begin{itemize}
\item $\AsiT$ is the true area of the i$^{th}$ cell in the source grid, that is the area used by the model on the source grid,
\item $\AdjT$, the true area of the j$^{th}$ cell in the destination grid,
\item $\AsiF$ is the estimated area of the i$^{th}$ cell in the source grid. The estimation is done and used inside SCRIP. We say that $\AsiF$ is the false area of the  i$^{th}$ grid cell, 
\item $\AdjF$, the estimated (false) area, of the j$^{th}$ cell in the destination grid,
\item $\Fsi$, the constant value of the field F in the i$^{th}$ cell of the source grid,
\item $\Fdj$, the regridded value of F in the j$^{th}$ cell of the destination grid. The coupler estimates $\Fdj$ from $\Fsi$,
\item $\Psi = \AsiT\times\Fsi$, integral of $\Fsi$ over the true area of the i$^{th}$ cell of the source map. It is an extensive variable,
\item $\Pdj = \AdjT\times\Fdj$, integral of $\Fdj$ over the true area of the j$^{th}$ cell of the destination map. It is an extensive variable,
\end{itemize}

From now on, we use 'SCRIP' as a shortcut of 'the conservative regridding option implemented in the SCRIP'.

\section{Definitions}

\begin{definition}\label{def:regridding}
A regridding scheme calculates the sparse matrix that transforms $\Fsi$ into $\Fdj$. The coefficients of the matrix are called $\fij$~:
\begin{equation}\label{eq:regridding}
\Fdj = \sum_i\fij\Fsi
\end{equation}
\end{definition}

\begin{definition}\label{def:regriddingPhi}
Alternatively, a regridding scheme can calculate the sparse matrix that transforms $\Psi$ into $\Pdj$. The coefficients of the matrix are called $\pij$~:
\begin{align}\label{eq:regriddingPhi}
\Pdj &= \sum_i\pij\Psi,\\
\intertext{$\pij$ derives from $\fij$~:}
\pij &= \fij\times\frac{\AdjT}{\AsiT}.
\end{align}
\end{definition}

\begin{definition}\label{def:conservative}
The regridding of F is globally conservative if and only if $\sum_i\Psi = \sum_j\Pdj$. From now on, we use 'conservative' in place of 'globally conservative'.
\end{definition}

\begin{definition}\label{def:locallyConservative}
The regridding of F from a source to a destination grid is locally conservative if and only if the flux on the destination grid cell is exactly equal to the sum of all the fluxes from the underlying source grid cells. If the intensive quantity $\Fsi$ is constant inside each source grid cell, this writes exactly
\beq
\Pdj = \sum_i[\Fsi\times(\AsiT\cap\AdjT)].
\eeq
This gives the two equivalent expressions for a locally conservative regridding scheme
\begin{align}\label{eq:locallyConservative}
\Pdj &= \sum_i[\Psi\times\frac{\AsiT\cap\AdjT}{\AsiT}]\\
\intertext{and}
\Fdj &= \sum_i[\Fsi\times\frac{\AsiT\cap\AdjT}{\AdjT}],\\
\intertext{which, given Eq.~\ref{eq:regridding}, writes}
\fij &= \frac{\AsiT\cap\AdjT}{\AdjT}.
\end{align}

\end{definition}{eq:regridding} 

Note that a locally conservative regridding scheme is also conservative (see the demonstration in Eq.~\ref{eq:identicalarea}).

\begin{definition}\label{def:noFluctuation}
The regridding doesn't introduce fluctuations if and only if a cons\-tant field is transformed into the same constant field. This can also be expressed as~: $\Fsi = F, \forall i$ implies $\Fdj = F, \forall i$, or, by using Eq.~\ref{eq:regridding}~:
\beq
\sum_i\fij = 1, \quad\forall j.
\eeq

\end{definition}

\section{Implicit hypotheses}\label{par:hypotheses}

In this report, we implicitly make the following assumptions over the source and the destination map~:

\begin{hypothesis}\label{hyp:true}
Each true destination grid cell is included in a subset of true source grid cells; the true grid cells are pairwise disjoint~: 
$$\cup_i\AsiT\supset\cup_j\AdjT$$ 
$$\cap_i\AsiT = \emptyset\quad\text{and}\quad\cap_i\AdjT = \emptyset$$
\end{hypothesis}
This implies that
\beq
\sum_i\AsiT\cap\AdjT = \AdjT
\eeq

\begin{hypothesis}\label{hyp:false}
Each false source grid cell is included in a subset of false destination grid cells; the false grid cells are pairwise disjoint~: 
$$\cup_i\AsiF\subset\cup_j\AdjF$$
$$\cap_i\AsiF = \emptyset\quad\text{and}\quad\cap_j\AdjF = \emptyset$$
\end{hypothesis}
This implies that:
\beq
\sum_j\AsiF\cap\AdjF = \AsiF
\eeq

\section{Analysis of current SCRIP}

The current version of SCRIP calculates $\fij$ (Def.~\ref{def:regridding}) as~:
\begin{equation}\label{eq:SCRIP}
\fijSCRIP = \frac{\AsiF\cap\AdjF}{\AdjF}
\end{equation}	
Note that, by definition, as for a locally conservative scheme (Def.~\ref{def:locallyConservative}),
\beq\label{eq:scripNoFluctuation}
\sum_i\fijSCRIP =  1,
\eeq
whereas, in general,
\beq
\sum_j\fijSCRIP \neq  1.
\eeq

Note also that, from Eq.~\ref{eq:scripNoFluctuation}, Def.~\ref{def:noFluctuation} and under Hyp.~\ref{hyp:false}, SCRIP regridding never introduces fluctuations. Under SCRIP, a constant field is always regridded into the same constant field.

\subsection{False and true grid areas are identical}

In this paragraph, we make the following hypothesis~:
\begin{hypothesis}\label{hyp:falseEqualTrueAreas}
The false and the true grid cell areas are identical, that is
\begin{align}
\AsiF &= \AsiT\label{eq:falseEqualTrueSource}\\ 
\intertext{and} 
\AdjF &= \AdjT\label{eq:falseEqualTrueDest}. 
\end{align}
This means that SCRIP is locally conservative (see Eq.~\ref{eq:SCRIP} and Def.~\ref{def:locallyConservative}).
\end{hypothesis}

Now, we show that Hyp.~\ref{hyp:falseEqualTrueAreas} implies that SCRIP is conservative~:
\beq
\begin{split}\label{eq:identicalarea}
\sum_j\Pdj &= \sum_j\Fdj\AdjT \\
&= \sum_j(\sum_i\fijSCRIP\Fsi)\times\AdjT \qquad\text{from Eq.~\ref{eq:regridding}}\\
&= \sum_i[\Fsi\times(\sum_j\fijSCRIP\AdjT)] \\
&= \sum_i[\Fsi\times(\sum_j\AsiT\cap\AdjT)] \qquad\text{from Eq.~\ref{eq:SCRIP} and Eq.~\ref{eq:falseEqualTrueSource}-\ref{eq:falseEqualTrueDest}}\\
&= \sum_i\Fsi\AsiT \qquad\text{from Hyp.~\ref{hyp:true}}\\
&= \sum_i\Psi.\\ 
\end{split}
\eeq

So, given Def.~\ref{def:conservative}, if the areas are estimated correctly, SCRIP is conservative.

\subsection{False and true grid areas are different}

\subsubsection{The coupler using SCRIP transmits $\Fsi$}

We now test if SCRIP is conservative when the false and the true grid areas are different~:
\beq
\begin{split}\label{eq:areadifferent}
\sum_j\Pdj &= \sum_i[\Fsi\times(\sum_j\fijSCRIP\AdjT)] \\
&= \sum_i[\Fsi\times\sum_j(\AsiF\cap\AdjF)\times\frac{\AdjT}{\AdjF}] \qquad\text{from Eq.~\ref{eq:SCRIP}}\\
&\neq  \sum_i\Fsi\AsiT.\\ 
\end{split}
\eeq
So, SCRIP is not conservative.

\subsubsection{The coupler transmits $\Psi=\Fsi\AsiT$ instead of $\Fsi$}

 We now consider the case in which the coupler transmits the extensive quantities $\Psi=\Fsi\AsiT$ instead of $\Fsi$. This means that we apply the following recipe~:
\begin{enumerate}
\item we don't modify the SCRIP code,
\item the source map sends $\Psi$ instead of $\Fsi$ to the destination map. 
\item in the destination map, we recover $\Fdj = \frac{\Pdj}{\AdjT}$ by dividing the output of the matrix multiplication with the true area of the destination grid cell. 
\end{enumerate}

Eq.~\ref{eq:regridding} becomes
\begin{align}\label{eq:scripflux}
\Pdj &= \sum_i\fijSCRIP\Psi,\\
\intertext{which also writes}
\Fdj\AdjT &= \sum_i\fijSCRIP\Fsi\AsiT,\\
\Fdj &= \sum_i[\fijSCRIP\times\frac{\AsiT}{\AdjT}\Fsi].
\end{align}

We want to determine if such a scheme makes SCRIP conservative~:
\beq
\begin{split}
\sum_j\Pdj &= \sum_j(\sum_i\fijSCRIP\Psi)\\
&= \sum_i\Psi(\sum_j\fijSCRIP) \\
\end{split}
\eeq
So,
\begin{alignat}{1}
\sum_j\Pdj &= \sum_i\Psi\\
\Leftrightarrow\qquad\sum_j\fijSCRIP &= 1 \qquad\forall i\\
\Leftrightarrow\qquad\sum_j\frac{\AsiF\cap\AdjF}{\AdjF} &= 1 \qquad\forall i\label{eq:conservativescripflux}
\end{alignat}

But, $\sum_j\frac{\AsiF\cap\AdjF}{\AdjF} \neq 1$ in the general case. So, transmitting $\Psi$ in place of $\Fsi$ doesn't render SCRIP conservative.

\section{Modifications needed to render SCRIP conservative}

We propose the modifications that need to be brought to SCRIP or to the use of SCRIP to make the regridding conservative in the general case.

\subsection{The coupler transmits the intensive quantities $\Fsi$}

We are looking for the new weights $\fijINT$ that would make SCRIP conservative, when the coupler transmits the intensive quantities $\Fsi$. 

From Eq.~\ref{eq:areadifferent}, SCRIP is conservative if and only if
\beq\label{eq:condfijint}
\sum_j\fijINT\AdjT = \AsiT\qquad\forall i
\eeq

We know that $\sum_j\frac{\AsiF\cap\AdjF}{\AsiF} = 1$ (see Hyp.~\ref{hyp:false}). So,
\beq\label{eq:fijint}
\fijINT = \frac{\AsiF\cap\AdjF}{\AsiF}\times\frac{\AsiT}{\AdjT}
\eeq
is a solution of Eq.~\ref{eq:condfijint}.

We can write $\fijINT$ differently~:
\beq
\begin{split}\label{eq:fijscripint}
\fijINT &= \fijSCRIP\times\frac{\AdjF}{\AsiF}\times\frac{\AsiT}{\AdjT}\\
&= \fijSCRIP\times\frac{\AsiT/\AsiF}{\AdjT/\AdjF}\\
\end{split}
\eeq

So, if the coupler using SCRIP is not modified (the $\Fsi$ are transmitted), SCRIP needs to know the true areas of the source and of the destination grid cells. We need to add a parameter (SCRIP\_\-CON\_\-SERVATIVE\_\-INTENSIVE) that tells SCRIP to compute $\fijINT$ (Eq.~\ref{eq:fijint}) in place of $\fijSCRIP$ (Eq.~\ref{eq:SCRIP}). 

This ''intensive'' modified SCRIP would always be conservative for intensive quantities.

\subsection{The coupler transmits the extensive quantities $\Psi=\Fsi\AsiT$}

Eq.~\ref{eq:conservativescripflux} would always be true if $\AsiF$ was the denominator in place of $\AdjF$.

We define
\beq
\begin{split}\label{eq:fijscripext}
f_{ij}^{SCRIP,ext} &= \frac{\AsiF\cap\AdjF}{\AsiF}\\
&= \fijSCRIP\times\frac{\AdjF}{\AsiF}\\
\end{split}
\eeq

If the coupler transmits the extensive quantities $\Psi$ instead of $\Fsi$, then SCRIP doesn't need to know the true grid areas. However, we need to add a parameter (for example SCRIP\_\-CONSERVATIVE\_\-EXTENSIVE) that tells SCRIP to compute 
$f_{ij}^{SCRIP,ext}$ (Eq.~\ref{eq:fijscripext}) in place of $\fijSCRIP$ (Eq.~\ref{eq:SCRIP}).

 This ''extensive'' modified SCRIP would always be conservative for extensive quantities.

\section{Discussion of the equations}\label{par:discussion}

We have demonstrated that SCRIP as such is not conservative in the general case. Transmitting extensive instead of intensive quantities between the two grids doesn't render SCRIP conservative. We have modified the interpolation scheme to guarantee conservation.

Having said this, we now discuss four points~, the first ones being specific to the climate community, the last one having a more general scope~:

\subsection{Effect of a global normalization}
When the grid cells are closer to the pole, the difference between true and false grid cell areas, as calculated by SCRIP, is larger. So, applying a global normalization would not resolve the imbalance between the pole and the equator introduced by the inaccurate conservative scheme. 

\subsection{Working on different sphere surfaces}
Sometimes, the models to be coupled work on different sphere surfaces. A global normalization is enough to correct for such a mismatch.

\subsection{Why an accurate conservative regridding scheme?}
 A small inaccuracy in the conservative regridding scheme is not random but it is both geographically localized and repeated at each time step of the simulation. This implies that it may alter significantly the result of the simulation and that the analysis of this accuracy is important.
 
\subsection{Numerical diffusion} 

For the first order conservative regridding scheme (the one treated here), the quantity is distributed from the source grid cell to all destination grid cells intersecting it (see for a comparison between first- and second-order remapping schemes the paper of Jones~\cite{jones_first-_1999}). The quantity is distributed not only to the portion of the destination grid cell intersecting the source grid cell, but also to the whole surface of this destination grid cell. It acts as an numerical diffusion of the quantity to be mapped. It is unavoidable in the case of a first-order remapping scheme.

When we modify the regridding scheme, we want to know if we increase this numerical diffusion.

In the current SCRIP, this diffusion is not determined by the intersections between the true source grid cells and the true destination grid cells (as it would be in the ideal case), but by the intersections between the false source grid cells and the false destination grid cells. If we change the equations of SCRIP as proposed in this report, the intersections are between the same grid cells. So, we will not significantly modify the extent of the numerical diffusion, which is good.

\subsection{Numerical fluctuations}

In SCRIP, $$\sum_i\fijSCRIP = 1.$$ So from Def.~\ref{def:noFluctuation}, the scheme doesn't introduce numerical fluctuations, that is a constant field is always transformed into the same constant field. But, $\sum_j\fijSCRIP \neq 1$, so, from Eq.~\ref{eq:areadifferent}, SCRIP is not conservative.

In the new scheme (SCRIP\_\-CONSERVATIVE\_\-INTENSIVE), $$\sum_j\fijINT = 1.$$ So, from Def.~\ref{def:locallyConservative}, it is locally and globally conservative. However, $$\sum_i\fijINT \neq 1.$$ So, from Def.~\ref{def:noFluctuation}, it introduces numerical fluctuations.

\subsection{Being conservative and avoiding numerical fluctuations}

Mathematically, in a first-order scheme, the scheme avoids numerical fluctuations if and only if the sparse matrix of the interpolation weights $[\fij]$ is a left stochastic matrix. The scheme is locally conservative if and only if $[\fij]$ is a right stochastic matrix. In both cases, $[\fij]$ is sparse.

The scheme has both properties if and only if $[\fij]$ is a sparse doubly stochastic matrix, that is a matrix for which each row and column sums to 1. We can state it as a constrained optimization problem and express it as Lagrange multipliers~: we want to find the sparse doubly stochastic matrix $[\fij]$ that is the closest to the matrix $[\fijSCRIP]$. The Lagrange function writes~:
\beq
\begin{cases}
L(\fij,\lambda_j,\mu_i) &= \sum_{ij}B_{ij}(\fij-\fijSCRIP)^2 \\
                        &+ \sum_j\lambda_j(\sum_i\fij-1) \\
                        &+ \sum_i\mu_i(\sum_j\fij-1) \\
\fij &= 0\quad \rm{if}\quad B_{ij} = 0
\end{cases}
\eeq

$B_{ij}$ is constant in this optimization problem. Its elements are set to force the sparsity of $[\fij]$. Ideally, we want the final matrix $[\fij]$ to have the same sparsity as the matrix $[\fijSCRIP]$. So, if the degrees of freedom of $[\fijSCRIP]$ are correctly distributed, the matrix $[B_{ij}]$ reads~:
\beq
B_{ij} = 
\begin{cases}
1& \rm{if}\quad \fijSCRIP \neq 0\\
0& \rm{if}\quad \fijSCRIP = 0
\end{cases}
\eeq
otherwise, we set more generally~:
\beq
\begin{cases}
B_{ij} = 1&\rm{if}\quad \fijSCRIP \neq 0\\
B_{ij} \leq 1 &\text{if\quad i close to j} \\
B_{ij} = 0 &\text{if\quad i far from j}  
\end{cases}
\eeq

Then, $[\fij]$ is the solution of the linear system~:
\beq
\begin{cases}
\fij = 0 &\text{if } B_{ij} = 0\\
\frac{\partial L}{\partial\fij} = 0 &\text{if } B_{ij} \neq 0\\
\frac{\partial L}{\partial \lambda_{j}} = 0 &\forall j\\
\frac{\partial L}{\partial \mu_{i}} = 0 &\forall i
\end{cases}
\eeq

The first-order interpolation using these interpolation weights is conservative and it avoids at the same time numerical fluctuations. As drawback, it can introduce a larger numerical diffusion.

\section{Conclusion}

The equations and the discussion presented above have a general scope. They apply whenever we want to make a first-order interpolation of a field between two grids, whose edges are not exactly known.

One way (the one used by SCRIP) is to estimate internally the edges and calculate the interpolation weights by directly computing the intersections of the estimated (false) areas of the source and of the destination grid cells. This method doesn't introduce numerical fluctuations but is not conservative.

We have proposed two simple modifications to this algorithm~:
\begin{enumerate}
\item if intensive quantities are used by the coupler, SCRIP needs to know the true grid cell areas of the source and of the destination map, and to modify the algorithm to calculate the weights $\fijINT$ (Eq.~\ref{eq:fijscripint}). This ''intensive'' modified SCRIP would always be conservative for intensive quantities.
\item if extensive quantities are used by the coupler, SCRIP needs to modify the algorithm to calculate the weights $f_{ij}^{SCRIP,ext}$ (Eq.~\ref{eq:fijscripext}). This ''extensive'' modified SCRIP would always be conservative for extensive quantities.
\end{enumerate}
Given the discussion of \S\ref{par:discussion}, we think that the climate community should apply these changes.\cite{chavas_conservative_2013}

\bibliography{conservativeRegridding}{}
\bibliographystyle{plain}
\end{document}